# Fluctuations of gas concentrations in three mineral springs of the East Eifel Volcanic field (EEVF)

G.M. Berberich[1], M.B. Berberich[2], A.M. Ellison[3]

[1]Technical University of Dortmund, Faculty of Electrical Engineering and Information Technology, Image Analysis Group, Otto-Hahn-Straße 4, 44227 Dortmund, Germany;
**Corresponding author**: gabriele.berberich@tu-dortmund.de; Tel: +49-231-755-4518 or +49-2235-955233

[2]IT-Consulting Berberich, Am Plexer 7, 50374 Erftstadt, Germany

[3]Harvard University, Harvard Forest, 324 North Main Street, Petersham, Massachusetts, 01366 USA

## Abstract

We present a geochemical dataset acquired during continual sampling over 7 months (bi-weekly) and 4 weeks (every 8 hours) in the Neuwied Basin, a part of the East Eifel Volcanic Field (EEVF). We used a combination of geochemical, geophysical, and statistical methods to identify potential causal processes underlying the correlations of degassing patterns of four gases (He, Rn, $CO_2$, $O_2$), earth tides, and tectonic processes in three mineral springs (Nette, Kärlich and Kobern). We explored whether temporal relations in gas concentrations in the three mineral springs could be indicators of hidden faults through which the gases migrate to the surface from deeper underground. Our results do not confirm $CO_2$ as a primary carrier gas for trace gases in all springs. Temporal analyses of the $CO_2$-He couple indicate that Nette and Kärlich are directly linked via a continuous tectonic fault in an ENE-WSW trending direction. There is also evidence that Kärlich and Kobern (NNE-SSW fault system) and Nette and Kobern (NW-SE fault system) are tectonically linked. These fault linkages are unknown previously but could be related to the rising numbers of earthquake events occurring in this area since 2010. We did not find any evidence that weather processes (e.g., barometric pressure), earth tides, or low local earthquake magnitudes actively modulate degassing. The volcanic activity in the EEVF is dormant but not extinct and to understand and monitor its magmatic and degassing systems, we recommend bi-weekly samplings at minimum.

## Key words

## Funding

The study is part of the research project "GeoBio-Interactions" funded by the VW-Stiftung (Az 93 403).

# 1    Introduction

The East Eifel Volcanic Field (EEVF; western Germany) is located on the left side of the still uplifting Rhenish Massif, which is part of the European Cenozoic Rift System. This part of the Rhenish Massif has been affected by complex major tectonic and magmatic processes. Today's uplift rates (≤3.5 mm/year; Campbell et al. 2002) are attributed mainly to plume-related thermal expansion of the mantle-lithosphere (Ritter et al. 2001, Walker et al. 2005, Tesauro et al. 2006), crustal thinning and associated volcanism (Clauser 2002), active rifting processes with motion discontinuities of 0.06–1.7 mm/year (Hinzen 2003), and possibly crustal-scale folding and/or the reactivation of Variscan thrust faults under the present-day NW–SE-directed compressional stress field (Hinzen 2003; Dèzes et al. 2004; Ahorner 1983, Ziegler and Dèzes 2005, Tesauro et al. 2006). The EEVF is characterized by intensive intra-continental Quaternary volcanism (100 volcanic eruption centres) and several alternating phases of volcanic activity and non-eruptive phases during the last 700 ka. The youngest eruption happed ≈12,900 years ago when the Laacher See volcano experienced a phreato-plinian eruption (Litt et al. 2001). The volcanic activity is dormant but not extinct and is certainly going to continue (Wörner 1998, Schmincke 2007).

Currently, NW-SE, N-S and ±E-W trending fault zones are highly-permeable, hydrogeological important structures, and provide permeable pathways for migration of $CO_2$-rich gases linked to Quaternary volcanic activities and rapid water circulation (May 2002, Clauser 2002, Campbell et al. 2002). In the EEVF and its adjoining Neuwied basin, more than 300 minerals springs and mofettes occur at faults and crosscut zones of regional tectonic lineaments, presumably merging into shear zones of the crust or upper mantle. Waters, enriched in crustal components (i.e., $Cl^-$, $Ca^{2+}$, or $SO_4^{2-}$), usually discharge as brines or, in case of rapid circulation rates, as thermal waters (Clauser 2002). Furthermore, $CO_2$-rich, mostly hydrogen-carbonate-waters, show significantly high concentrations of crustal- or upper mantle-derived chemical tracers, such as He and Rn (Griesshaber 1998, Clauser et al. 2002, Bräuer et al. 2013). The highest concentrations of mantle-derived $^3He$ are around Laacher See with $^3He/^4He$ ratios are between 1.3 and 5.6 $R_a$. Additionally, ratios of $Mg^{2+}/Ca^{2+}$ between 0.75 and 1 in mineral water chemistry indicate mantle influence (Clauser 2002).

Though many investigations (volcanological, petrochemical, and petrological) have been carried out in the EEVF in the last 40 years, data collection and monitoring has been only annual (e.g. Bräuer et al. 2013), and no continual long-term measurements or analyses of geological gas concentrations in mineral springs has been done. Here we present new insights into concentration changes and time series of carbon dioxide ($CO_2$), helium (He), radon (Rn) and oxygen ($O_2$) in three minerals springs in the Neuwied Basin over a period of 7 months within the research project "GeoBio-Interactions" (March – September, 2016).

The most reliable tracer gases are $CO_2$ and noble gases such as He, and Rn (Boubron et al. 2002, Wilkinson et al. 2010, Gilfillan 2011). $CO_2$ serves as carrier gas and is a confirmed fault indicator (Ciotoli et al. 2004). $CO_2$ is a major gas in soils and groundwater, and mainly produced by biologic activity and equilibrium with carbonate minerals. However, part of the $CO_2$ can also originate from mantle degassing or metamorphic processes (Giroud et al. 2012). He is considered an ideal geochemical tracer of crustal fluid movement (Ciotoli et al. 2004) and has two naturally-occurring radiogenic isotopes: radiogenic He ($^4$He) and mantle derived He ($^3$He). Radiogenic He is always produced in the crust due to radiogenic decay of radioactive isotopes such as $^{238}$U, $^{235}$U and $^{232}$Th and may dilute $^3$He in relation to its vertical flow rate (Kennedy et al. 1997). The mantle is the source of He (primordial origin). In the atmosphere, He concentration is considered homogenous (at 5.2204 ±0.0041ppm; Davidson & Emerson 1990) and reflects an approximate balance between the supply from degassing of the solid Earth and thermal escape from the top of the atmosphere into space (Kockarts 1973; Sano 2010). Because of its high mobility, Rn (half-life 3.82 days), a direct product of $^{226}$Ra in the $^{238}$U decay series, can be used as an additional tracer that provides a qualitative measure of gas migration (Voltattorni et al. 2010; Ma et al. 2012).

We used a combination of geochemical, geophysical, and statistical methods to identify potential causal processes underlying the correlations of degassing patterns of the four aforementioned gases, earth tides, and tectonic processes in three mineral springs. We explored whether temporal relationships among gas concentrations in the three mineral springs could be indicators of hidden faults for migrating geogenic gases from the deep underground. Specifically, we tested the null hypotheses that concentrations of the investigated geogenic gases were independent in all three springs. Additionally, temporal associations between the fluctuations and variations in gas concentrations in all three mineral springs were investigated to test whether there was a simultaneous effect and what influence weather, earth tides, and seismic activity could have had on degassing.

## 2    Methods

We monitored water chemistry, concentrations of $CO_2$, He, Rn and $O_2$, time series of carrier-trace-gas couples ($CO_2$-He and $CO_2$-Rn) and variability of fluctuations in gas concentrations during a 7-month bi-weekly sampling campaign ('7-M'; 1 March–30 September 2016) and during one month every 8 hours sampling ('4-W'; 12 July–11 August 2016).

### 2.1    Study area

The triangle-shaped study site formed by the three investigated mineral springs was situated in the seismically active Neuwied Basin, a part of the EEVF, near Koblenz, Germany (Fig. 1a). The square-shaped Neuwied Basin (≈585 km$^2$) is a tectonic depression at the crosscut zone of the Mosel and Rhine rivers through with a maximum depth of 350 m which evolution started in the Eocene and is still persisting (Meyer 1994). The Paleozoic basement beneath the study site consists of up to 5 km alternating strata of Lower Devonian, e.g. shales, sandstones, Greywacke, clay, and siltstones ("Hunsrückschiefer" and "Singhofener Schichten"). Paleogene clay layers (≈ 70 m) which contribute to the major part of the basin fillings and are turning into a more calcareous and marine type at depth, Quaternary volcanites (pumice and trass) of the Laacher See eruption (≈13.000 years ago), and Loess sediments are covering the Devonian units (Meyer 1994, LGB-RLP 2005; LGB-RLP 2017b). Faults in the Neuwied Basin are trending NE-SW (Variscan direction) and NW-SE (present-day stress field). Their ages and total penetration depths are mostly unknown. Earthquakes (Fig.1) concentrating in the seismically active "Ochtendunger Fault Zone (OFZ)" (Ahorner 1983) and that are related to stress-field-controlled block movements showing a weak-to-moderate seismicity occur mostly in a shallow crustal depth (≤15 km) with local magnitudes (Richter scale) rarely exceeding 4.0.

The three investigated mineral springs Flöcksmühle in the Nette river near Ochtendung (hereafter: 'Nette'), Waldmühle in Mülheim-Kärlich (hereafter: 'Kärlich') and 'Kobern' in Kobern-Gondorf are located West (Nette), East (Kärlich) and in the South (Kobern) of the Ochtendung Fault Zone (Ahorner 1983).

Nette (Fig. 1b) is a cold-water geyser, well-bored to a depth of 100m in 1928 for the extraction of carbonic acid. The geyser's plumbing geometry consisted of a vertical well-bore which allowed for the upward artesian migration of $CO_2$-rich fluids. The positive feedback system of a $CO_2$-driven eruption occurred within the well, leading to a soaring height of 3-4 m. After selling the rights for carbonic acid production in 1939, the geyser was shut. The owner opened the geyser again in 1965, but without using it commercially. In 1967, the geyser was plugged by vandals with debris and its regular $CO_2$-driven eruption cycles faded out, while a rhythmically $CO_2$-rich fluid discharge of 120 L/min continued. To prevent further damage to the spring, the outlet was re-shaped into its present-day U-form. In 1999, the former restrictions from the sell-off of exploitation rights expired (pers. comm. J. Dumont, 2016).

Kärlich (Fig. 1c), well-bored in 1978, is located approx. 15 m south of the Lützel creek (Lützelbach). The freely available mineral water is collected by the inhabitants and used as drinking water.

Kobern (Fig. 1d), with a discharge of 1.5 L/min, is located approx. 1 km North of Kobern-Gondorf in a mixed forest stand next to the Hohenstein creek (Hohensteinbach). It was presumably constructed in its present-day shape at the beginning of the 20th century (Liesenfeld 1926). Kobern's outer shape is constructed as a cave. Its mineral water is collected in a semicircular stonewalled water basin. The $CO_2$-rich discharge is entering the basin from the bottom. A casing record is not available.

All three sampled mineral springs are located at fault intersections: Nette and Kobern are located close to the centre of the OFZ; Kärlich approx. 7 km away (Fig. 1a).

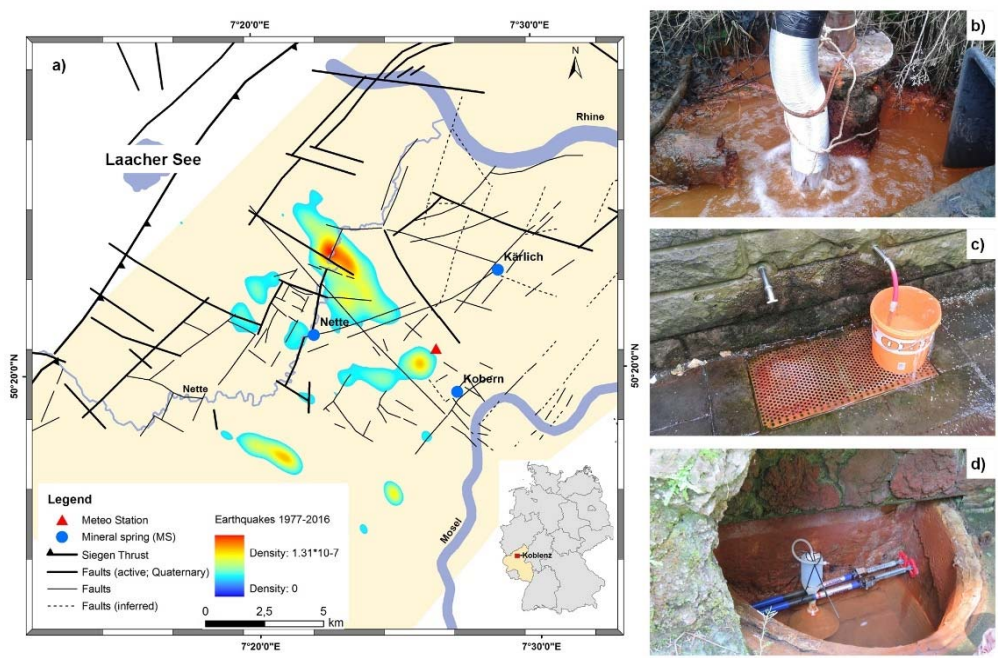

Fig. 1    Location of the three mineral springs (blue dots) ≈15 km SE of the Laacher See volcano within the Neuwied Basin (light yellow area). The map (a) shows tectonic structures (black lines) and active faults (bold black lines according to Meyer 1994) compiled from different tectonic maps, location of the meteorological station at the Goloring site (red triangle), and probability density of the earthquake events from 1977-2016 which are related to the Ochtendunger Fault Zone (rainbow contours). $CO_2$, He, Rn, and $O_2$ were sampled from three mineral springs. The inset shows the location of study site within Germany. Photographs show (b) Nette and (c) Kärlich with modified outlets, and (d) Kobern equipped with a permanent gas mouse (all photographs: G. Berberich).

## 2.2    Gas sampling and geochemical analyses

We combined different analyses of mineral spring gas and mineral waters in a field campaign from 1 March – 30 September 2016 (7 months; '7-M'; 16 samplings), to investigate biweekly and, during one month (4-weeks campaign ('4-W'), every eight hours; 79 samplings), relationships between

- $CO_2$, He, Rn, $O_2$, and H2S concentrations in mineral springs,
- geophysical processes (seismic events, earth-tides) and
- meteorological conditions.

During the survey, a total number of 1968 samples were collected. Gas sampling followed the procedure described in Berberich (2010). The gas phase of the three mineral springs was sampled according to the sampling pattern. To facilitate gas sampling, the outlet of the Nette was modified so that gas sampling could be performed without sampling atmosphere (Fig. 1b). At Kärlich (Fig. 1c) a U-shaped copper pipe was docked to the outlet and the gas was sampled beneath the water surface in a basket using a mobile gas mouse. At Kobern (Fig. 1d) a permanent gas mouse was installed. The mineral waters also were analysed to investigate their major ion contents and water provinces.

Helium$_{(total)}$ samples were transferred into a 20-ml stainless-steel "gas frog" and were analyzed immediately after collection, by a mobile, modified mass spectrometer (Alcatel ASM 142; adixen) which had been converted to a 20-ml sample volume for a single He measurement. Rn samples were transferred into a 100-ml Lucas cell and analysed after a resting period of three hours minimum using a Lucas detector (JP048; Radon Detector LUK4). To obtain a quantitative measurement of $CO_2$, H2S and $O_2$ concentrations, a portable Dräger-meter equipped with three sensors with different detection limits was docked to the probe or the gas mouse and run at every sampling for two minutes (Dräger X-am ® 7000; DrägerSensor® Smart IR $CO_2$ HC, measuring range 0–100% by volume, DrägerSensor® Smart XS EC H2S 100 ppm; range 0 - 100 ppm $H_2S$ and DrägerSensor® XS 2-$O_2$, range 0–25 Vol %). Details of analytical procedures and errors were described in Berberich (2010). Water temperature and pH value were measured using the WTW 320 pH-combined electrode with integrated temperature sensor SenTix 97 T (pH value measuring range: -2.0–16.0, resolution: 0.01; temperature measuring range: -5 – 99.9°C, resolution: 0.1K). Electric conductivity was measured using Greisinger electronic device GLM 020A (measuring range: from 0 µS/cm to 19.99 mS/cm).

## 2.3    Meteorological parameters

Meteorological data were obtained from a radio meteorological station (WH1080) placed 2 m above ground at the Goloring site (which was part of the research project "GeoBio-Interaction", Fig. 1a) in the center of the triangle-shaped study area and in ≈ 6 km distance from the springs. It continuously logged meteorological conditions (temperature [°C], humidity [%], air pressure [hpa], wind speed [m/s], rainfall [mm], dew point [°C]) at 5-min intervals. The recorded data were downloaded every two days, checked for completeness, and stored in a data base. Meteorological data were used to investigate the potential influence of weather conditions on the fluctuation of the gas concentrations in mineral waters.

## 2.4    Earthquake events

Data on earthquake events for the sampling campaign were collected from the seismological databases provided by the Erdbebenstation Bensberg (BNS 2017) (www.seismo.uni-koeln.de/events/index.htm) and by the Landesamt für Geologie und Bergbau, Rheinland-Pfalz (LGB-RLP 2017c) (http://www.lgb-rlp.de/fachthemen-des-amtes/landeserdbebendienst-rheinland-pfalz/).

## 2.5    Earth tides

Cyclic changes in the earth's environment are caused by the gravitational pull of both the Sun and the Moon on the earth. The results are two slight lunar and two solar tidal bulges (earth tides). The two bulges occur at the surface of the earth that approximately faces the Moon and at the opposite side while the Earth rotates around its axis. Earth tides were calculated using the tool developed by Dehant et al. & D. Milbert version 15.02.2016 (http://geodesyworld.github.io/SOFTS/solid.htm). To investigate whether degassing concentrations were delayed relative to earth tides, we used cross-correlations and the approach of Crockett at al. 2006 to the 4-W sampling.

## 2.6 Data analysis

All analyses were done using R version 3.3.2 (R Core Team 2016) or MATLAB R2017a. We examined the effects of six meteorological variables on geogenic gas concentrations: Air temperature (°C), barometric pressure (hPa), dew-point (°C), relative humidity (%), hourly rainfall (mm), and windspeed (km/h). As many of these variables are correlated with one another, we used principal components analysis (R function prcomp) on centred and scaled data to create composite "weather" variables (i.e., principal axes) that were used in subsequent analyses. We used the "median+2MAD" method (Reimann et al. 2005) to separate true peaks in geogenic gas concentrations from background or naturally elevated concentrations: any observation greater than the overall median+2MAD was considered to be anomalous. For interpreting the significance of the correlation coefficient we followed Hinkle et al. (2009). Fluctuation patterns were investigated by centering and scaling of the data and removing the outliers (i.e., points > 1.5 standard deviations from entered and scaled data were excluded from analysis (7-M: ≈10%; 4-W: ≈9%). Then the smoothing spline method, was applied, where f(x) was piecewise polynomial computed from smoothing parameter p (p = 0.9999) where x is normalized by mean and std. Cross-correlation analyses were applied to investigate temporal relations between degassing patterns of sampled springs and carrier-tracer gas relations. Factor analysis of sampled geogases and chemical parameters were carried out, to define elements with similar behaviors and close relations.

## 3 Results

### 3.1 Water chemistry

The waters of Nette, Kärlich and Kobern are $CO_2$-gas/bicarbonate-rich mineral waters of Ca-Mg-$HCO_3^-$-type (Fig. 2) and characterized by low discharge temperatures (8.5 – 17 °C), weakly acidic pH (pH 5.2 – 6.5), high $Mg^{2+}$ (mean 130 mg/L), $Ca^+$ (mean 106 mg/L); $HCO_3^-$ (≈ 2400 mg/L), and low $SO_4^{2-}$ (<100 mg/L) and $Cl^-$ (<50 mg/L) concentrations. Mg/Ca ratios were 0.8 for all springs. Water temperatures in all three springs varied only slightly during 7-M and 4-W sampling and were moderately to highly correlated among springs (Table 1; Annex 1). The concentration of salts, such as Na+, K+, Cl- and $SO_4^{2-}$ was twice as high and four time higher respectively, in Nette (904 mmol/L) relative to Kobern (423 mmol/L) and Kärlich (219 mmol/L).

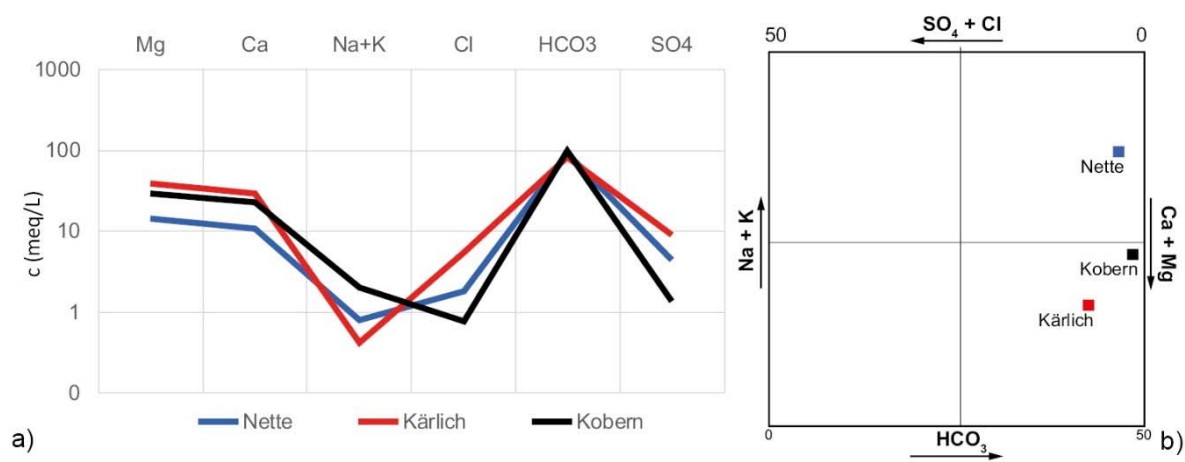

**Fig. 2**      **Schoeller (a) and Langlier (b) diagram showing major ion contents and water provinces of the three mineral waters of Nette, Kärlich and Kobern**

## 3.2    Gas composition

$CO_2$ is the major component of the investigated gases relative to He, Rn and $O_2$. No H2S was recorded in the springs. Median $CO_2$ values were different for all springs: very high concentrations (86 Vol.% – 92 Vol.%) were found in Nette and Kärlich during 7-M and 4-W both samplings. Kobern always had median concentrations < 80 Vol.% (Table 1).

He concentrations (Table 1) were highly variable during both samplings and fluctuated over an order of magnitude, e.g. Nette between 20 and 110 ppm (4-W). These concentrations were 10-fold higher than at Kärlich (6 -10 ppm He) or Kobern (1 – 7 ppm He). Almost 84% of the He concentrations in Kobern were below atmospheric standard (5.22 ppm; Davidson & Emerson 1990).

Rn concentrations also varied a lot (Table 1) with the highest values in Kärlich (114 Bq/L) and the lowest in Nette (≈3 Bq/L). Median concentrations of $O_2$ in Nette and Kärlich were very low but similar (≈1 Vol.%.); Kobern had the highest $O_2$ concentration (≈ 5 Vol.%).

**Table 1**      **Descriptive statistics of gas concentrations in the three mineral springs for the (a) 7 months and (b) 4-weeks-campaign**

| | | 7-M (bi-weekly; 01.03. – 30.09.2016) | | | | | a) | 4-W (8-hrs. 12.07. – 11.08.2016) | | | | | b) |
|---|---|---|---|---|---|---|---|---|---|---|---|---|---|
| | | N | Mean | Median | Min | Max | SD | Median +2MAD | N | Mean | Median | Min | Max | SD | Median +2MAD |
| **Nette** | CO₂ (Vol.%) | 16 | 89.88 | 90.00 | 82.00 | 96.00 | 3.96 | 96.28 | 76 | 84.79 | 88.00 | 62.00 | 94.00 | 6.34 | 97.27 |
| | He (ppm) | 16 | 47.06 | 49.74 | 17.37 | 58.17 | 10.91 | 63.63 | 79 | 49.20 | 50.29 | 20.83 | 110.40 | 13.24 | 67.55 |
| | Rn (Bq/L) | 16 | 6.47 | 6.26 | 2.62 | 9.46 | 1.88 | 9.11 | 79 | 8.01 | 6.58 | 0.56 | 59.90 | 8.08 | 12.92 |
| | O₂ (Vol.%) | 16 | 1.12 | 0.95 | 0.70 | 2.60 | 0.62 | 1.82 | 76 | 1.79 | 1.05 | 0.70 | 7.60 | 1.60 | 3.44 |
| | Temp. (°C) | 16 | 14.33 | 14.50 | 12.20 | 15.70 | 0.94 | 15.91 | 79 | 14.55 | 14.60 | 12.90 | 17.20 | 0.68 | 15.59 |
| | pH | 15 | 6.23 | 6.26 | 5.66 | 6.45 | 0.18 | 6.49 | 79 | 6.23 | 6.25 | 5.72 | 6.42 | 0.10 | 6.38 |
| | Cond. (mS/m) | 14 | 3.13 | 3.14 | 1.22 | 4.86 | 0.78 | 4.12 | 79 | 2.93 | 2.85 | 1.68 | 4.22 | 0.31 | 3.22 |
| **Kärlich** | CO₂ (Vol.%) | 16 | 91.07 | 92.00 | 86.00 | 96.00 | 3.61 | 98.26 | 76 | 86.34 | 86.00 | 74.00 | 96.00 | 3.95 | 91.91 |
| | He (ppm) | 16 | 7.77 | 7.80 | 6.21 | 9.03 | 839.47 | 9.16 | 79 | 7.84 | 7.77 | 6.02 | 10.95 | 0.82 | 9.06 |
| | Rn (Bq/L) | 16 | 72.44 | 73.69 | 38.32 | 114.02 | 15.22 | 92.00 | 79 | 73.73 | 78.11 | 4.91 | 92.33 | 15.58 | 99.71 |
| | O₂ (Vol.%) | 16 | 0.91 | 0.90 | 0.10 | 1.50 | 0.30 | 1.30 | 76 | 1.16 | 1.10 | 0.70 | 3.70 | 0.48 | 1.74 |
| | Temp. (°C) | 16 | 12.34 | 12.45 | 10.60 | 14.10 | 1.10 | 14.25 | 79 | 12.55 | 12.50 | 10.00 | 15.40 | 0.88 | 13.75 |
| | pH | 15 | 5.95 | 5.92 | 5.22 | 6.81 | 0.31 | 6.28 | 79 | 5.83 | 5.81 | 5.68 | 6.50 | 0.11 | 5.95 |
| | Cond. (mS/m) | 14 | 1.79 | 1.74 | 1.13 | 3.21 | 0.46 | 2.24 | 79 | 1.72 | 1.64 | 1.28 | 3.90 | 0.34 | 1.98 |
| **Kobern** | CO₂ (Vol.%) | 16 | 78.13 | 79.00 | 50.00 | 92.00 | 10.42 | 94.25 | 76 | 74.61 | 74.00 | 64.00 | 88.00 | 4.49 | 81.34 |
| | He (ppm) | 16 | 2.97 | 2.02 | 0.98 | 1.11 | 2.55 | 5.56 | 79 | 4.38 | 4.48 | 1.00 | 7.50 | 1.24 | 6.23 |
| | Rn (Bq/L) | 16 | 45.78 | 46.63 | 16.90 | 64.84 | 11.41 | 62.48 | 79 | 48.66 | 49.97 | 7.82 | 79.78 | 9.67 | 61.47 |
| | O₂ (Vol.%) | 16 | 4.54 | 5.10 | 0.90 | 10.20 | 2.50 | 8.71 | 76 | 5.17 | 5.30 | 1.10 | 8.20 | 1.33 | 7.34 |
| | Temp. (°C) | 16 | 11.18 | 11.50 | 8.50 | 13.10 | 1.53 | 14.03 | 79 | 12.06 | 12.00 | 10.40 | 14.40 | 0.70 | 13.10 |
| | pH | 15 | 6.20 | 6.27 | 5.37 | 6.43 | 0.25 | 6.56 | 79 | 6.28 | 6.27 | 6.12 | 6.53 | 0.07 | 6.38 |
| | Cond. (mS/m) | 14 | 3.03 | 2.60 | 2.32 | 5.39 | 1.01 | 3.96 | 79 | 2.40 | 2.35 | 1.12 | 2.84 | 0.21 | 2.60 |

### 3.3 Factor analysis

Factor analysis aimed at defining elements with similar behaviors and close relations were applied to the results for the 7-M and 4-W samplings. Factor analyses for the 7-M sampling is unspecific (Fig. 3a-c). Distribution diagrams of factor loads indicate clear relations between $CO_2$ and He for the 4-W sampling for Nette and Kärlich, but not for Kobern. $CO_2$ seems not to be the main carrier for Rn in the springs (Fig. 3d-f). Also, $O_2$ cannot be grouped.

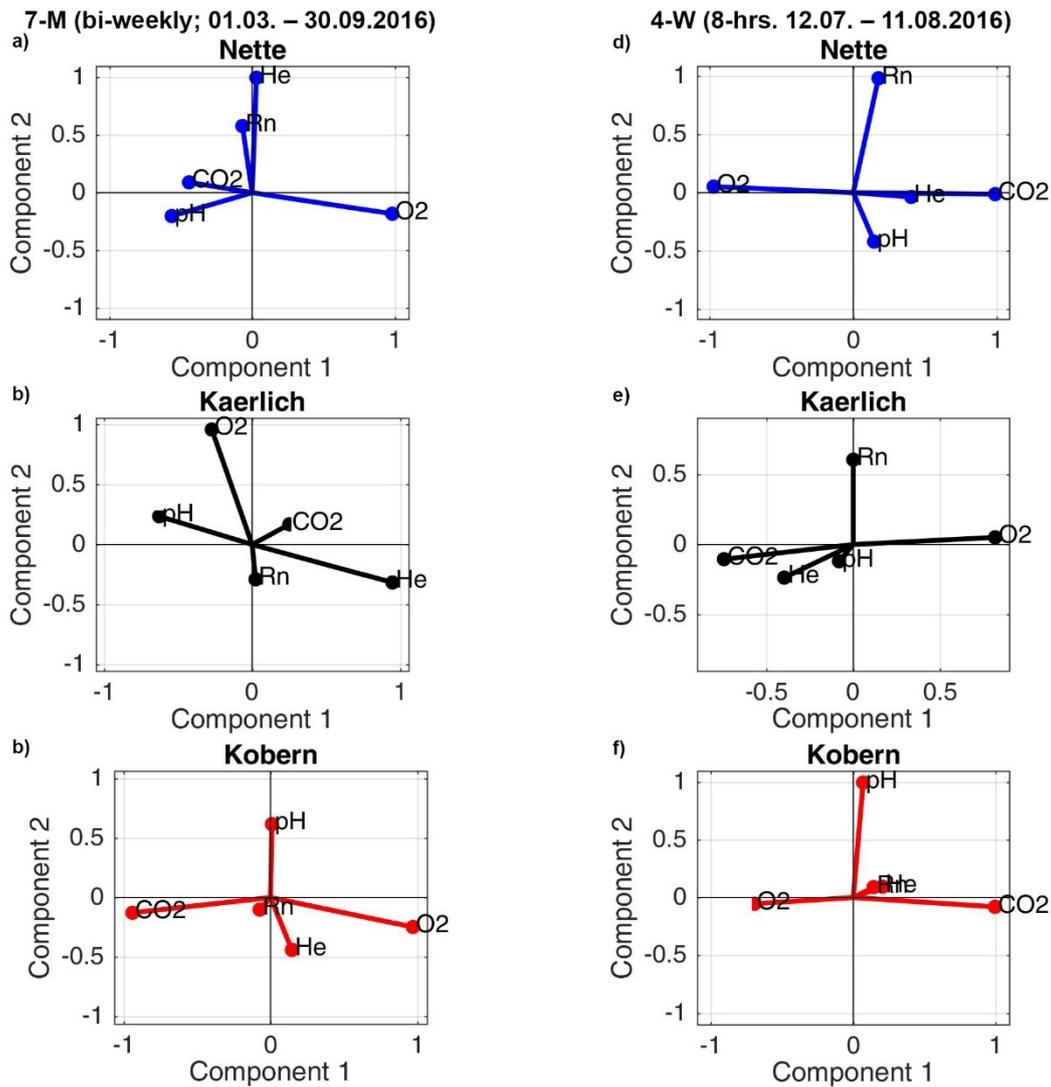

**Fig. 3**        **Factor 1–Factor 2 diagrams for the (a) bi-weekly and (b) 8-hours sampling in Nette, Kärlich, and Kobern**

## 3.4    Time series

### 3.4.1   Fluctuations in gas concentrations

A comparison of random samplings of gases between 2009 and 2014 (Fig. 4) with the mean concentration of 2016 revealed annual fluctuations. $CO_2$ concentrations were > 80 Vol.% in all springs and only slightly variable. $CO_2$ concentrations in Kobern were more variable (Fig. 4a). Concentrations for He declined in Nette and Kärlich, but rose slightly in Kobern between 2009 and 2015 (Fig. 4b). Rn concentrations were much more variable: declining in Nette and Kobern but stable at high levels in Kärlich (Fig. 4c). Mean He and Rn concentrations for Nette and Kobern in 2016 exceeded the mean of previous samplings. In the gas phase of Kärlich, the parameter limit for Rn (100 Bq/L) of German Drinking Water Ordinance (TrinkwV 2015) was exceeded only once during the 7-M sampling campaign.

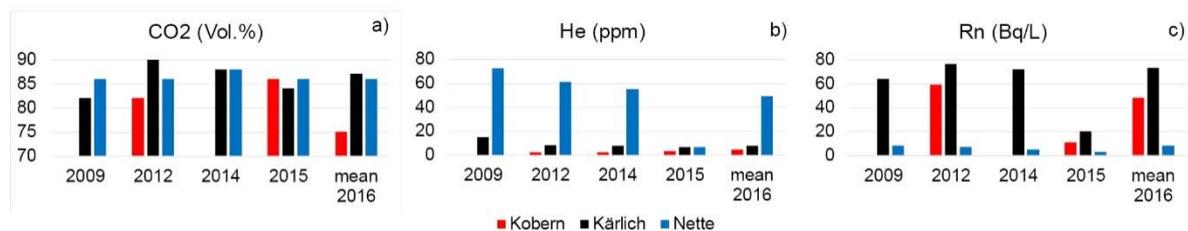

**Fig. 4** **Geogas concentrations in three mineral springs between 2009 and 2016. Average concentrations for 2009 – 2015 are given as red lines and numbers modified after (Schreiber & Berberich 2012)**

During the 2016 survey, gas fluctuation patterns could be observed for the 7-M and the 4-W sampling. During the bi-weekly sampling, concentrations varied by an order of magnitude, e.g. Kärlich between 38 and 114 Bq/L for Rn (Table 1). Sinusoidal gas fluctuations were higher in amplitude than during the 4-W campaign (Fig. 5). Variations in $CO_2$ and He were highly correlated (0.92 and 0.75) in Nette and Kärlich (Fig. 5a,b,d; Annex 3). He fluctuations in Kobern (-0.59 and -0.53; Fig. 5b) were anticyclic relative to Nette and Kärlich. Rn fluctuations were not as pronounced as the other gases (Fig. 5c) and only little correlated between Kobern and Kärlich. $O_2$ fluctuations were variable in all springs. There were moderate to very high correlations between Rn fluctuations and weather variables (temperature and air pressure) in Kobern and Kärlich (Fig. 5c): Temperature (-0.71) was negatively correlated with Rn in Kobern, whereas air pressure was correlated anticyclically (-0.66) in Kärlich. In Kobern, $CO_2$ fluctuation patterns were positively correlated with wind speed, but negatively correlated with temperature; for $O_2$ the correlations were *vice versa* (Annex 3).

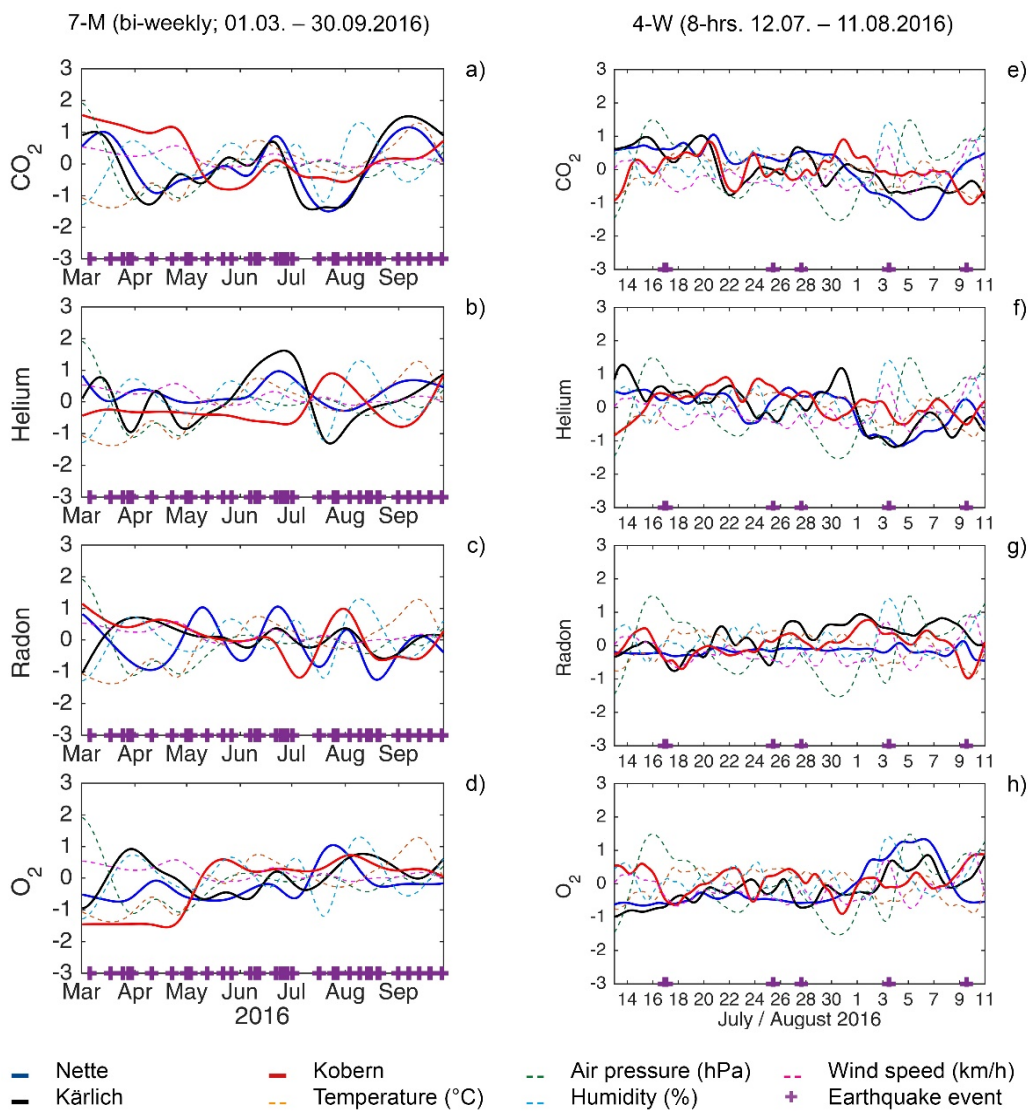

**Fig. 5** Fluctuation patterns of geogas concentrations (centered and scaled data) during the bi-weekly (a-d) and 8-hours (e-h) survey in the three mineral springs and weather conditions (temperature = orange; atmospheric pressure = green; humidity = cyan; wind speed = magenta) and earthquake events (purple crosses)

Fluctuations during the 4-W campaign (Fig. 5e-h; Annex 3) showed similar patterns but were more detailed than those seen in the 7-M sampling. Fluctuations were similar for $CO_2$, He and $O_2$ (Fig. 4 e,f,h) in Nette and Kärlich. There was a moderate positive correlation between Rn fluctuations in Kärlich and Kobern. Moderate wind speed negatively influenced while temperature positively influenced $CO_2$ fluctuation patterns in Kobern (Fig. 4f). Temperature moderately influenced He fluctuations in Kobern (Fig 4g). In all three springs, $CO_2$ and He concentrations declined towards the end of the 4-W campaign (Fig. 4e,f). There were moderate fluctuation patterns for $O_2$ in Kärlich and Nette (Fig. 4 g,h). Gas fluctuations in Kobern almost always differed from Nette and Kärlich (Fig. 4e-h).

### 3.4.2 Temporal variations of concentrations and carrier-trace gas couples in springs

Cross-correlations for $CO_2$ (Fig. 6a) and He (Fig. 6b) suggested that Nette and Kärlich are directly and instantaneously linked with respect to these gases at an approximate lag of 2 days. Cross-correlations for Nette and Kobern, and Kobern and Kärlich revealed a positive but low relationship with a time lag of ≈2 days for $CO_2$ (Fig. 6a). For Rn, a small positive relationship with a lag of 8 hours was observed between Nette and Kobern, whereas the relationship was small and negative with a lag of ≈80 hours between Nette and Kärlich (Fig. 6c). Kobern and Kärlich were directly and instantaneously linked with respect to Rn (Fig. 6c).

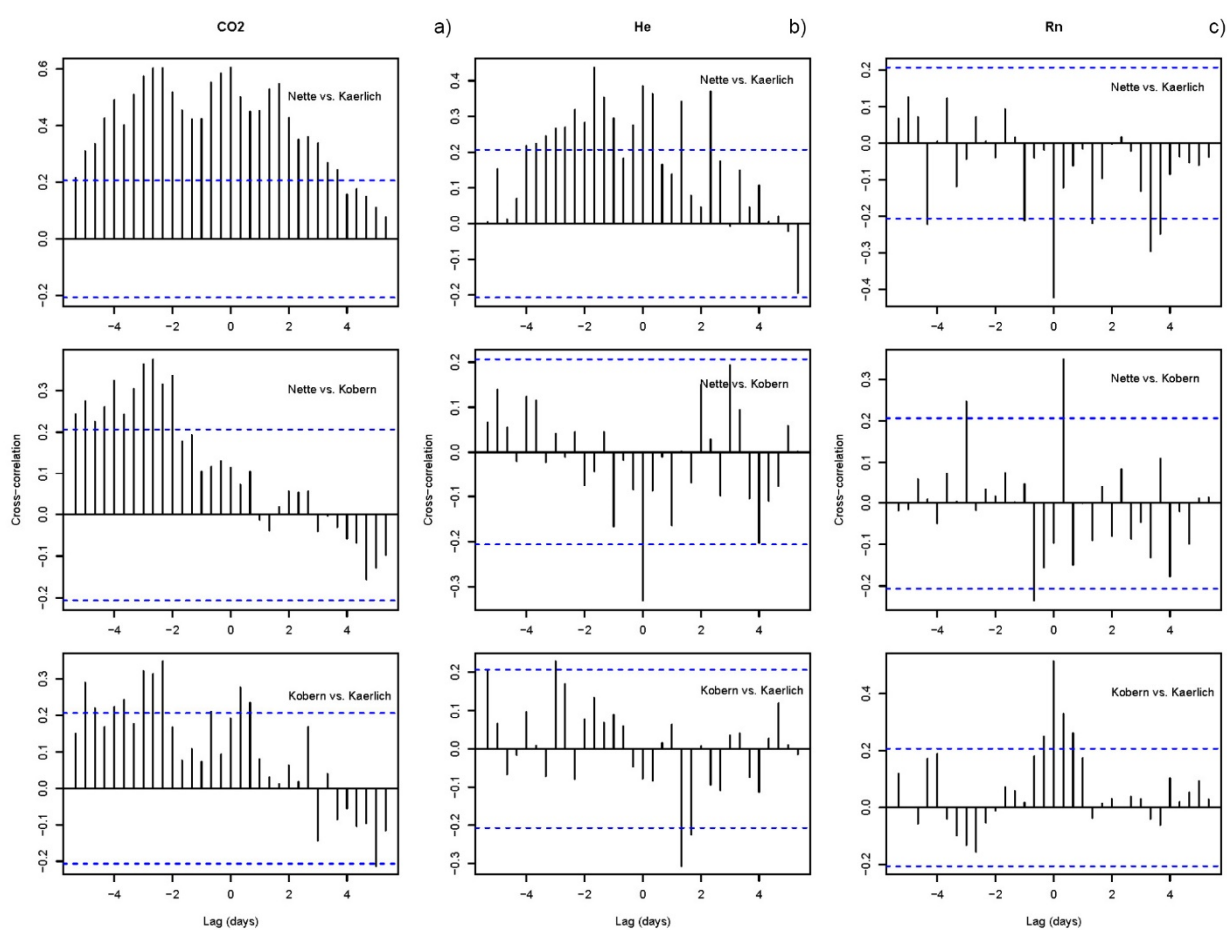

**Fig. 6** Cross-correlation of the time-series of (a) $CO_2$, (b) He and (c) Rn (8-hours; median smoothed) between all three springs for the 4-W campaign. Blue dashed lines indicate confidence thresholds

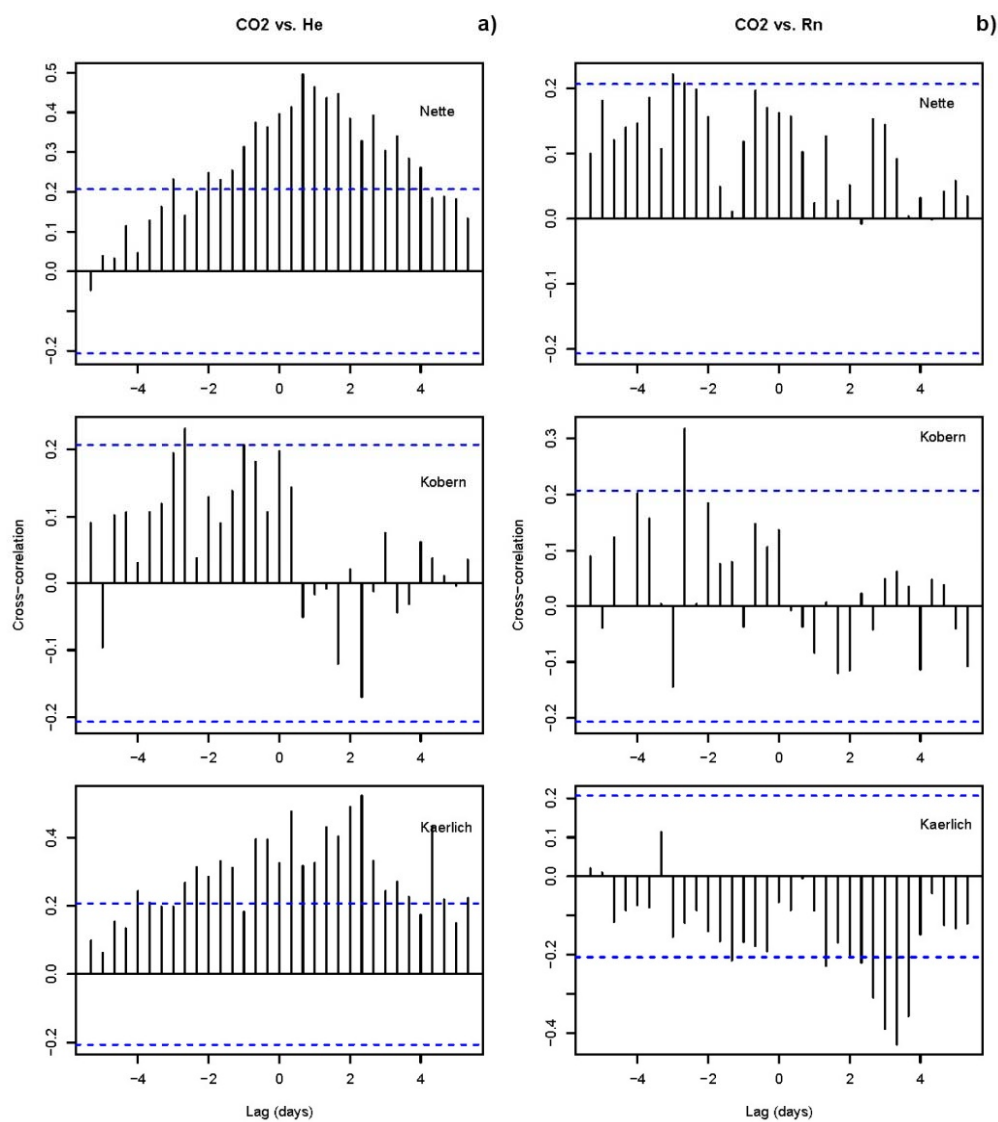

**Fig. 7**      **Cross-correlation of the time-series of (a) $CO_2$ vs. He and (b) $CO_2$ vs. Rn (8-hour median smoothed) in all three springs for the 4-W campaign. Blue dashed lines indicate confidence thresholds**

Joint visualization of the time series of carrier-trace gas couples reveal differences among all springs. A coupled system of $CO_2$-He is visible in Nette (lag = 16 hours) and Kärlich (lag = 56 hours). The cross-correlation for the $CO_2$-He couple in Kobern never exceeded 0.25 (Fig. 7a). In Nette no $CO_2$-Rn coupled system was visible. In Kobern there was a positive but low relationship with a positive time lag of about ≈3 days for Rn, in Kärlich a low negative relationship with a negative time lag of ≈ 80 hours (Fig. 7b).

## 3.5    External factors

### 3.5.1    Meteorological conditions

Stable meteorological conditions with only small variations were recorded during the 7-M and 4-W campaign. Air temperatures ranged from -0.6–33.5 °C (mean = 14.6 °C), with a single maximum rainfall event (66.9 mm) on 13 March. Variation in atmospheric pressure and wind speed were small (Table 2). The degassing processes from mineral springs did not appear to be influenced by meteorological conditions (Table 2 and 3, Annex 2a, b).

**Table 2**      **Meteorological conditions during the (a) 7-months (March – September 2016) and (b) 4-week field campaign (12 July - 11 August 2016); SD = Standard deviation, CV = Coefficient of Variation**

|  | 7-M (bi-weekly; 01.03. – 30.09.2016)   a) | | | | | | 4-W (8-hrs. 12.07. – 11.08.2016)   b) | | | | | |
|---|---|---|---|---|---|---|---|---|---|---|---|---|
|  | Min | Max | Mean | Median | SD | CV | Min | Max | Mean | Median | SD | CV |
| Temperature (°C) | -0.60 | 33.50 | 14.62 | 15.10 | 6.13 | 0.42 | 5.70 | 33.50 | 18.03 | 17.70 | 4.38 | 0.24 |
| Atm. Pressure (hPa) | 965.20 | 998.30 | 981.88 | 982.30 | 6.06 | 0.01 | 978.20 | 994.20 | 984.98 | 984.80 | 3.80 | 0.00 |
| Dew-point (°C) | -6.30 | 22.30 | 9.27 | 10.60 | 5.43 | 0.59 | 5.00 | 21.40 | 13.15 | 13.40 | 3.04 | 0.23 |
| Rel. humidity (%) | 20.00 | 99.00 | 72.70 | 74.00 | 17.66 | 0.24 | 30.00 | 99.00 | 75.15 | 77.00 | 16.36 | 0.22 |
| Rainfall (mm) | 0.00 | 66.90 | 0.05 | 0.00 | 0.65 | 13.32 | 0.00 | 3.30 | 0.03 | 0.00 | 0.17 | 6.06 |
| Windspeed (km/h) | 0.00 | 23.40 | 2.09 | 1.10 | 2.12 | 1.02 | 0.00 | 12.20 | 1.48 | 1.10 | 1.71 | 1.16 |

**Table 3**      **Results of PCA on weather variables**

|  | PC1 | PC2 | PC3 |
|---|---|---|---|
| **Loadings (> |0.3|)** | | | |
| Air pressure | | | -0.59 |
| Temperature | -0.68 | | |
| Dew Point | -0.51 | 0.34 | 0.45 |
| Relative Humidity | 0.34 | 0.64 | |
| Rainfall | | | 0.50 |
| Windspeed | | -0.57 | 0.31 |
| **Cumulative Variance Explained (%)** | **31** | **56** | **79** |

### 3.5.2    Earthquakes

In 2016, the number of earthquake events in the study area was twice as high (82 events) as the mean per year (45) between 1977-2015. These earthquakes were lower in local magnitude (ML 0.4) but deeper in depths (≈8 km; Table 4). During the 7-M sampling, 43 small-scale earthquakes (ML -0.7 – 1.8; depth: 1 – 26.7 km) occurred, while five occurred during the 4-W sampling (ML -0.1 – 1.4; depth: 3 – 11.5 km) (Fig. 8a). One earthquake (24 June; ML 0.7) occurred at the lower crust ≈ 27 km depth (Fig. 8b). Two additional ones, on 9 August, were ≈24 km away. Because of their far distance and low magnitudes ($M_L$ 0.8) these earthquakes were discarded from further analyses. No correlations between earthquakes and gas concentrations were observed for the 7-M campaign. During the 4-W sampling, a decline in $CO_2$ and He concentrations and a rise of Rn and $O_2$ concentrations could be visually observed after local two local earthquakes on 25 July (ML 1.4; Depth 10 km) and 27 July (ML 0.7; depth 9 km) at distances less than 10 km to the three springs (Fig. 5e,f).

**Table 4**    **Mean number of earthquake events, local magnitude (ML) and depth (km) for 1977-2015 compared to 2016 events**

| Year | No. | ML (mean) | Depth (km; mean) |
|------|-----|-----------|------------------|
| 1977-2015 | 45 (mean per year) | 1.0 | 6.1 |
| 2016 | 82 | 0.4 | 8.2 |

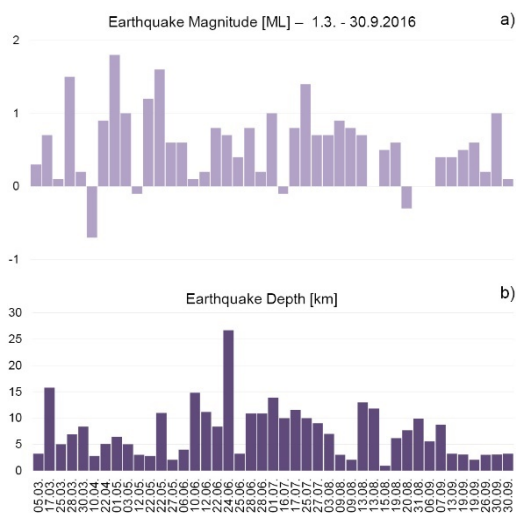

**Fig. 8**    **Local magnitudes (a) and depths (b) of earthquake events during the sampling period**

### 3.5.3   Earth tides

Earth tides were basically semi-diurnal. We observed a positive but low influence of earth tides on $CO_2$ (0.40) and He (0.32) fluctuation patterns during the 4-W campaign at Kärlich. Rn fluctuations in Nette were anticyclic (-0.43) to earth tides (Annex 4). No correlations between earth tides and gas concentrations were observed in Kobern. Cross-correlation of lagged earth tides and gas concentrations showed little if any correlation in any of the springs.

# 4    Discussion

## 4.1    Water chemistry

In general, the three analyzed $CO_2$-driven mineral waters in the Neuwied Basin share a similar geochemical composition apart from their salt content. Their geochemical compositions are comparable to the ones of the Eifel region published by Griesshaber (1998), LGB-RLP (2012) and Bräuer et al. (2013). They are representatives of deep end-member magmatic cold weakly acidic mineral waters, which are extremely enriched in $CO_2$-gas and mantle-derived fluids. These fluids can be identified due to their unique chemical characteristics, e.g. by high $Mg^{2+}/Ca^{2+}$-ratios. Large ratios of Mg/Ca and Cl/C indicate strong mantle or crustal signatures. Each of the ratios sufficiently characterizes these signatures by themselves (Clauser et al. 2002). Dissolution of volcanic material by $CO_2$-gas enriched waters leads to $Mg^{2+}$-enriched waters (Clauser et al. 2002). The high $Mg^{2+}/Ca^{2+}$-ratios (0.8) found in all sampled springs can be attributed according to mantle components (Clauser et al. 2002).

The salt concentration in Nette is 2-4 times higher than in the other two springs, so there has to be an additional source of salts within the subterranean catchment area of this spring. This could be explained by two scenarios. First, a fault zone or a fracture network provides pathways for a brine that may directly reach the well. Nette is located at a crosscut zone of two fault systems close to the seismically active OFZ (Fig. 1), and there is a high probability of Nette penetrating a set of fracture networks within this fault system that may be collecting more brine rich fluids. Second, the Paleogene clay layers are showing a marine influence at depth (Meyer 1994). As Nette was well-bored to a depth of 100 m, it may reach the marine layers which may also provide salts for the brine. During the both sampling periods, the fluid pH slightly decreased when electrical conductivity increased. This process further highlights the coupled nature of $CO_2$ and brine that migrates upward during discharge process.

## 4.2    Gas composition

Continual bi-weekly or 8-hour sampling intervals revealed robust data for the three mineral springs. Prior analyses in the Eifel volcanic field were based on only 6 to 15 samples in total per location, which were gained on different, annual samples (e.g., Bräuer et al 2013). Bi-weekly samplings interval showed a much higher variability in concentrations and fluctuation patterns than the 4-W sampling in the three springs. Variability is even higher if gases are only sampled once a year. Furthermore, our results show that also daily fluctuations have to be taken into account when analyzing gas compositions and fluctuation patterns of time series, e.g. the decline of gas concentrations towards the end of the 4-W sampling (Fig. 5e-h) would not have been observed by larger sampling intervals. Presently the volcanic activity in the Eifel region is dormant but not extinct; longer non-eruptive phases alternate with volcanically active periods (Schmincke 2007). However, no volcanic monitoring system is installed at EEVF. To understand and monitor the EEVF's magmatic and degassing system in relation to new developments in earthquake processes, large sampling intervals are not useful and may lead to wrong assumptions. Therefore, we recommend using at least bi-weekly samples for a minimum of five years.

All sampled $CO_2$-driven springs are located at fault intersections and characterized by cool waters (10-17 °C) bubbling continuously and forming small red to orange-colored travertine deposits at the surface due to the presence of hematite ($Fe_2O_3$) and iron hydroxide ($Fe[OH]_2$) (Pfanz 2008). The sampled springs are characterized by a geyser-type (Nette) or mofette-type (Kärlich and Kobern) degassing systems, indicating fluid transport along high permeable migration paths. Though not anymore actively soaring up, Nette's degassing patterns can still be addressed as a $CO_2$-driven cold-water geyser-like system conducted by a man-made well-bore. This can be observed by Nette's more or less rhythmically fluid discharge. According to Watson (2014), there are four steps leading to an eruption within the wellbore: i) pressure reduction of $CO_2$-rich fluids and evolving of $CO_2$ gas; ii) exceedance of the surrounding fluid pressure by internal $CO_2(aq)$ pressure and start of nucleation, growth and merging of $CO_2$-bubbles; iii) hydrostatic pressure reduction resulting from increasing $CO_2$ bubble volume fraction and expansion of $CO_2$ bubbles; and iv) surface discharge. $CO_2$-$H_2O$ leakage through a well-bore is initiated by large pressure differences between the formation and the well (Watson 2014). Artesian conditions resulting in formation overpressure may be caused at Nette by the Paleocene clay layers (thickness up to 70 m) basin fillings (Meyer 1994). Furthermore, formation permeability governs the supply rate of $CO_2$-rich water (Watson 2014). If Nette is located in a seismically active area at a fault intersection of an active fault (Meyer 1994), permeability may be provided by tectonic processes. In contrast, Kärlich and Kobern can be characterized by a discharge of $CO_2$-rich fluids without any rhythmically periodicity. The mineral water of Kärlich is continually discharging, sparkling water with very small $CO_2$ bubbles in the fluid. At Kobern, the degassing pattern is slower. Larger $CO_2$ bubbles are migrating towards the surface mostly within minutes and are released into stagnant water being collected in a small stone-walled water basin.

$CO_2$ is a major gas in soils and groundwater, and mainly produced by biologic activity and equilibrium with carbonate minerals. It also serves as carrier gas and is a confirmed fault indicator (Ciotoli et al. 2004). However, part of the $CO_2$ can also originate from mantle degassing or metamorphic processes (Giroud et al. 2012). $CO_2$ concentrations in the gas phases can be divided into two sections: high concentrations in Nette and Kärlich and much lower ones in Kobern. The relation of Nette and Kärlich with regard to $CO_2$ may be explained by both springs being located within the degassing center of the EEVF (Bräuer et al. 2013) and on a same NE-SW trending seismically active fault system. This is not the case for Kobern which is located south of the center and on another fault system (Fig. 1). Furthermore, similar $CO_2$ concentrations in Nette and Kärlich suggesting that they are supplied by the same magmatic reservoir in the subcontinental mantle (Bräuer et al. 2013).

The great differences in He concentrations and in He/$CO_2$ ratios (Nette: 571, Kärlich: 90, Kobern: 55) in the three springs may be also attributed to their different tectonic locations. High He fluxes are often found in tectonically active zones, where seismic activity tends to maintain a high permeability in active fractures. These fractures act as preferential conduits for gases trapped in the mantle or the crust, leading to gas concentration anomalies (Baubron et al., 2002). Judging from the high He concentration in Nette, the transport must take place in highly-permeable conduits. This can be confirmed because Nette is located at a seismically active crosscut zone of two fault systems. In combination with its geyser-like $CO_2$-driven system, where gas phases reach very quickly the surface, the less soluble but highly volatile He is enriched in the free gas phase. Furthermore, Nette's high He concentrations can be attributed to geochemical tracers of crustal fluid movement (Ciotoli 2004) in this area with the mantle or radiogenic decay as sources. Kärlich, located on a fault system intersection that is not seismically active (Meyer (1994), had a He/$CO_2$ ratio twice as high as Kobern. In the former, the fracture network might be more permeable and provide degassing pathways. Kobern is a mofette-type spring with stagnant waters, and the chemical fractionation of gases tends to zero in such waters (Bräuer et al. 2013). Compared to the both other springs, Kobern is probably located on a less developed fault system, so that there is less permeability for fluid and degassing. Nevertheless, Nette and Kobern (NW-SE fault system) also seem to be tectonically linked, which might be related to the rising numbers of earthquake events occurring in this area since 2010 (BNS 2017). In the EEVF, the Variscan NE-SW direction is well known by the "Siegen Thrust" south of Laacher See volcano and NE-SW trending faults in the Neuwied Basin that was tectonically re-activated during the Quaternary (Meyer 1994).

The highest Rn concentrations were found in Kärlich, located at a fault intersection. In the study area, local highly elevated Rn potential (> 100 Bq/L) is associated with tectonic fault zones and clefts caused by advective gas transport along faults between the interbedding layers of Lower Devonian clay and siltstones as bedrocks and the Cenozoic sediment basin fillings (Kemski et al. 2012; LGB-RLP 2017b). The moderate positive correlation of Rn concentrations between Kärlich and Kobern, together with their temporal cross-correlation, suggest that these two springs might be directly linked with respect to Radon (Annex 2b). This might be evidence either for the availability of a common underground radon source or a heretofore unknown fault system connecting both springs in NNE-SSW direction. The positive, albeit small, correlation in Rn concentrations between Kärlich and Nette (Annex 2b, Annex 3) provides additional evidence that both springs could be connected by the ENE-WSW trending fault system (Fig. 1). Based on our results, we suggest this fault system to be linked between both springs, rather than interrupted as it shown on geologic/tectonic maps so far (Fig. 1).

We also note that Rn occurrence in drinking water is controlled by, e.g., underlying geology, tectonics, hydrological processes influencing groundwater, and domestic uses of water (Crocket et al. 2006). The Council Directive 2013/51/Euratom issued by the European Commission defines requirements for the concentrations of radioactive substances in water intended for human consumption. Member States had to specify a parametric value for radon between 100 and 1000 Bq/L, which should not be exceeded. In Germany, this parametric value is fixed in the drinking water Guideline (TrinkwV 2016) at 100 Bq/L. During the sampling campaign, there was no evidence that this value in the drinking water of Kärlich was permanently exceeded. Nevertheless, Rn values are constantly high (Table 1) and because local inhabitants collect mineral water from this spring for daily consumption, the drinking water should be monitored on a regular base to prevent negative health impacts, especially on small children.

## 4.3 Temporal variations of concentrations and carrier-trace gas couples in springs

Griesshaber (2000) and Clauser (2002) suggested $CO_2$ as primary carrier phase in the Eifel fluid-rock system where high concentration of $CO_2$ gas stimulates dissolution and precipitation processes. However, our results do not confirm $CO_2$ as a primary carrier gas for trace gases in all springs (Fig. 6 and 7). Rather, we could only couple $CO_2$ and He between Nette and Kärlich, (Fig. 5 and 7).

He concentrations are very low in Kobern, below the atmospheric standard at 5.22 ppm (Davidson & Emerson 1990). He in Kobern might be transported by another carrier gas, such as $N_2$, as suggested by Hong et al. (2010) for active faults in Taiwan. Bräuer et al. (2013) also suggested $N_2$ as carrier gas, but only for the periphery along the Rhine. Kobern also might be part of the $N_2$ periphery area and this area may be extending into the Neuwied Basin. Future investigations throughout the Neuwied Basin should investigate the $N_2$-He carrier-trace gas couple.

Radon anomalies at the earth's surface are caused by Rn migration via conduits or faults (Etiope and Martinelli 2002). Because of its short half-life (3.82 days) and limited migration, Rn either is transported rapidly towards the surface or there is a $^{226}Ra/^{238}U$ source in local aquifers and/or in the weathered bedrock (Padilla et al. 2013). Advection is the most important transport mechanism for Rn migration at the scale of tens to hundreds of meters (Kemski et al. 2012). The gas ascent is governed by geo-physical properties (e.g., variations in temperature, pressure, mechanical stresses, chemical reactions and mineral precipitation, porosity, micro-fractures or voids in bedrock) and hydrological parameters (e.g., flow direction of groundwater) (Kemski et al. 2012, Padilla et al. 2013). Rn will mix with other gases in the rocks to form a gas mixture (Crocket et al. 2006). The $CO_2$-Rn couple is considered to be the most probable carrier-gas mechanism (Baubron et al., 2002; Ciotoli et al. 2004). In our study area, however, evidence for $CO_2$-Rn couples could be identified only for Kärlich and Kobern but not for Nette, although the Lower Devonian siltstone (grain size < 2 mm) has a high emanation rate of approx. 25 % (Kemski et al 2012). This might be attributed to the fact, that the bedrock at Nette and consists of Devonian "Hunsrückschiefer" with more quarzitic components (LGB RLP 2005). The conditions in Nette are comparable to the Wallenborn coldwater geyser in the Westeifel Volcanic Field with comparably low Rn concentrations but high $CO_2$ concentrations (Berberich 2010). High dynamics in both geyser systems might prevent higher Rn concentrations and coupled systems. At Kärlich, the lag of 3.3 days is very close to the half-life of Rn. The low negative relationship can be explained by a decline in Rn emanation followed by a higher $CO_2$ degassing. The low positive relationship in Kobern might indicate a coupled system within the half-life of Rn.

## 4.4    External factors

### 4.4.1    Meteorological influence

Degassing pattern from the subsurface have been suggested to be dependent on meteorological conditions (e.g. Hinkle 1994; Toutain and Baubron 1999, Gal et al. 2011). Our results, based on continuously *in-situ* monitoring of the nearby meteorological variables do not support this suggestion (see also Padilla et al. 2013 for the El Hierro Volcanic system). Additionally, our results do not support the hypothesis that temperature is a dominant controller of $CO_2$ production (cf. Baubien et al. 2014). The reason for this difference might be attributed to the fact that the meteorological conditions in previous publications were not monitored continuously, but based on only one daily value recorded from distant meteorological stations (*cf*. Gal et al. 2011, Baubien et al 2014). The negative moderate influence of temperature and positive influences of wind speed in Kobern during the bi-weekly sampling can be attributed to the permanently wooden coverage of the semicircular stonewalled water basin that is in place for hazard prevention. This influence vanished during the 4-W sampling when the stonewalled water basin was ventilated every eight hours.

### 4.4.2    Earth tides

Deformations of the earth's crust by earth tides also have been suggested to influence gas concentrations (e.g., Rn; Barnet et al. 1997, Crocket et al. 2006). Earth tides are associated with cyclic variations in water-table level within the rock strata (Crocket et al. 2006), but the influence of earth tides on gas fluctuation patterns was low only during the 4-W sampling. The absence of a strong effect of earth tides on Rn could be because i) the emanating layer is located too deeply and any pumped Rn is diluted during migration (Crocket et al. 2006); ii) the study area is too far away from coastlines, ameliorating influences of earth tides; or iii) the 8-hour sampling interval was too large to capture effects of semi-diurnal earth tides.

### 4.4.3   Earthquake events

The behaviour of trace gas transport is governed by advection induced by pressure gradient and diffusion caused by concentration gradients. Increase in compressive stress, changes in the volume of the pore fluid or rock matrix, and fluid movement or buoyancy are important mechanisms driving fluid flow and keeping fractures open as pressure is higher than minimum principal stress (e.g., Birdsell et al. 2015, Boothroyd et al. 2016). During the study period in 2016, 43 earthquakes occurred in the EEVF. At least 25% of these occurred at depths > 10 km, and one (24 June: ML 0.7) was even deeper ($\approx$ 27 km depth). This is a new development in the EEVF. The deep depths of the hypocenters are clearly above average for the tectonic earthquakes within the Ochtendunger Fault Zone and can be related to magmatic processes at the lower crust beneath the EEVF and Neuwied Basin (LGB-RLP 2017a). Though such processes cannot be excluded, no relation between earthquakes events and the high gas concentrations were observed based on the bi-weekly 7-M samples. This result is attributable to either: (a) the earthquakes were too small to influence gas concentrations; (b) most of the earthquakes occurred >10 km from the springs to have any influence (Fig 13 a-c); (c) sampling intervals missed the occurrence of the small earthquakes.

On the other hand, evidence of seismic influence on fluctuation patterns could be visually observed after the two nearby (< 10 km) earthquakes that occurred on 25 July (ML 1.4; Depth 10 km) and 27 July (ML 0.7; depth 9 km) during the 4-W campaign (Fig. 13 e-g). Increases in Rn and $O_2$, and decreases in $CO_2$ and He concentrations were observed $\approx$ 1 day after each of these earthquakes. As there were only two such nearby events, however, statistical power is much too low to detect their relationships with gas fluxes. The recent occurrences of deep earthquakes that are related to magmatic processes at the lower crust suggests continuous monitoring in this youngest volcanic field in Germany. Such data also would help assess relationships between gas flux dynamics and earthquake events in the Neuwied Basin.

# 5  Conclusion

Geochemical composition of water and gas from three mineral springs in the Neuwied Basin, a part of the East Eifel Volcanic Field (EEVF), were sampled to determine water chemistry, gas composition, fluctuation patterns, and temporal variations between tracers of releases of natural geogenic gases (He, Rn, $CO_2$, $O_2$). Results of the continuous sampling during 7 months (bi-weekly) and 4 weeks (every 8-hours) are:

- The deep end-member magmatic cold, weakly acidic mineral waters share a similar geochemical composition. They are extremely enriched in $CO_2$-gas and mantle-derived components. The higher salt content in the Nette spring highlights the coupled nature of $CO_2$ and brine that migrates upward during discharges.

- Our results confirmed that $CO_2$ is the primary carrier gas for He in the Nette and Kärlich springs, but not in the Kobern spring. In the latter, the carrier gas for He may be $N_2$. Kobern also may be part of the $N_2$ periphery area along the Rhine; this area may be extending into the Neuwied Basin.

- Temporal analyses of the $CO_2$-He couple indicate that Nette and Kärlich are directly linked via a continuous tectonic fault trending ENE-WSW. There is also evidence that Kärlich and Kobern (NNE-SSW fault system) and Nette and Kobern (NW-SE fault system) are tectonically linked. Both fault linkages were previously unknown.

- We did not find any evidence that weather variables, such as barometric pressure or earth tides, actively modulate degassing.

- Small earthquakes occurred >10 km from the studied mineral springs but had virtually no influence on gas fluctuations.

- The volcanic activity in the EEVF is dormant but not extinct. To understand and monitor its magmatic and degassing systems in relation to new developments in earthquake processes, we recommend continuous bi-weekly sampling.

## Acknowledgements

We thank Hayley Simpson, Alfredo Román Tejeda and Stanley Obamwonyi (all three are Master Students at the University of Duisburg-Essen), and Mark Schumann, Felix Dacheneder and Thomas Evert (University of Duisburg-Essen) for gas sampling and analyses. Gas analyses were done using equipment from the department of Geology at University of Duisburg-Essen. We also thank Michael Dötsch (Major of Kobern-Gondorf), Jürgen Dumont (private owner of the Flöcksmühle (Nette) in the Nette river) and Uli Klöckner (Major of Mülheim-Kärlich) for their support and the permission to use and modify the outlet of the mineral springs Nette, Kobern and Kärlich for the field campaign.

The study is part of the research project "GeoBio-Interactions" funded by the VW-Stiftung (grant number Az 93 403) within the initiative "Experiment!" – Auf der Suche nach gewagten Forschungsideen.

## Annexes

**Annex 1a-b** Correlation coefficients for weather-PCA and water chemistry parameters (electric conductivity (µS/m), pH value, and temperature (°C)) for the three mineral waters of Nette, Kärlich and Kobern for (a) the 7-M and (b) the 4-W surveys

a)

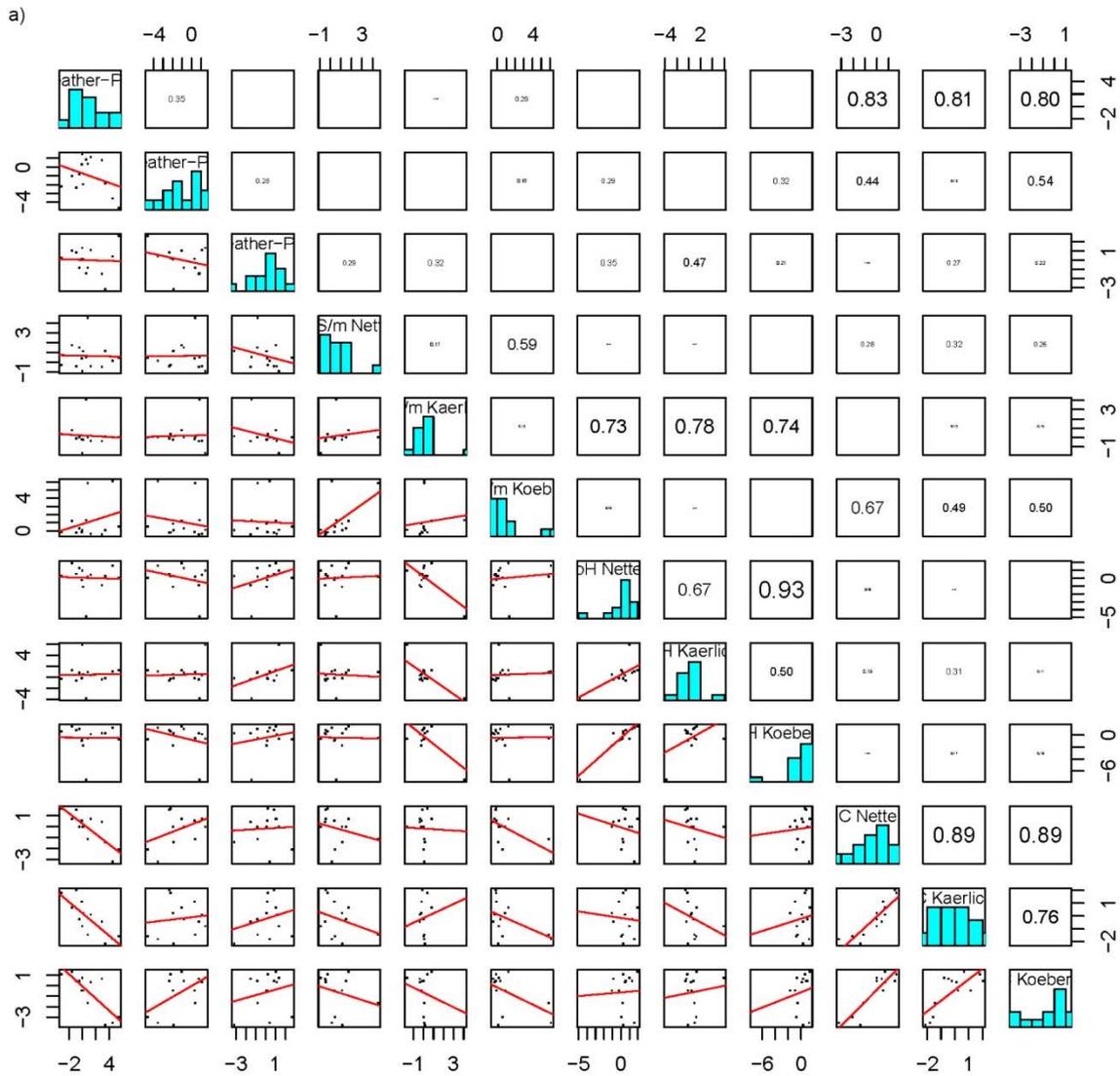

b)

**Annex 2a    Correlation coefficients of the principal components analysis of weather variables with geogenic gas concentrations for (a) $CO_2$, (b) He, (c) Rn and (d) $O_2$ for the three mineral springs during the 7-M campaign**

7-M (bi-weekly; 01.03. – 30.09.2016)

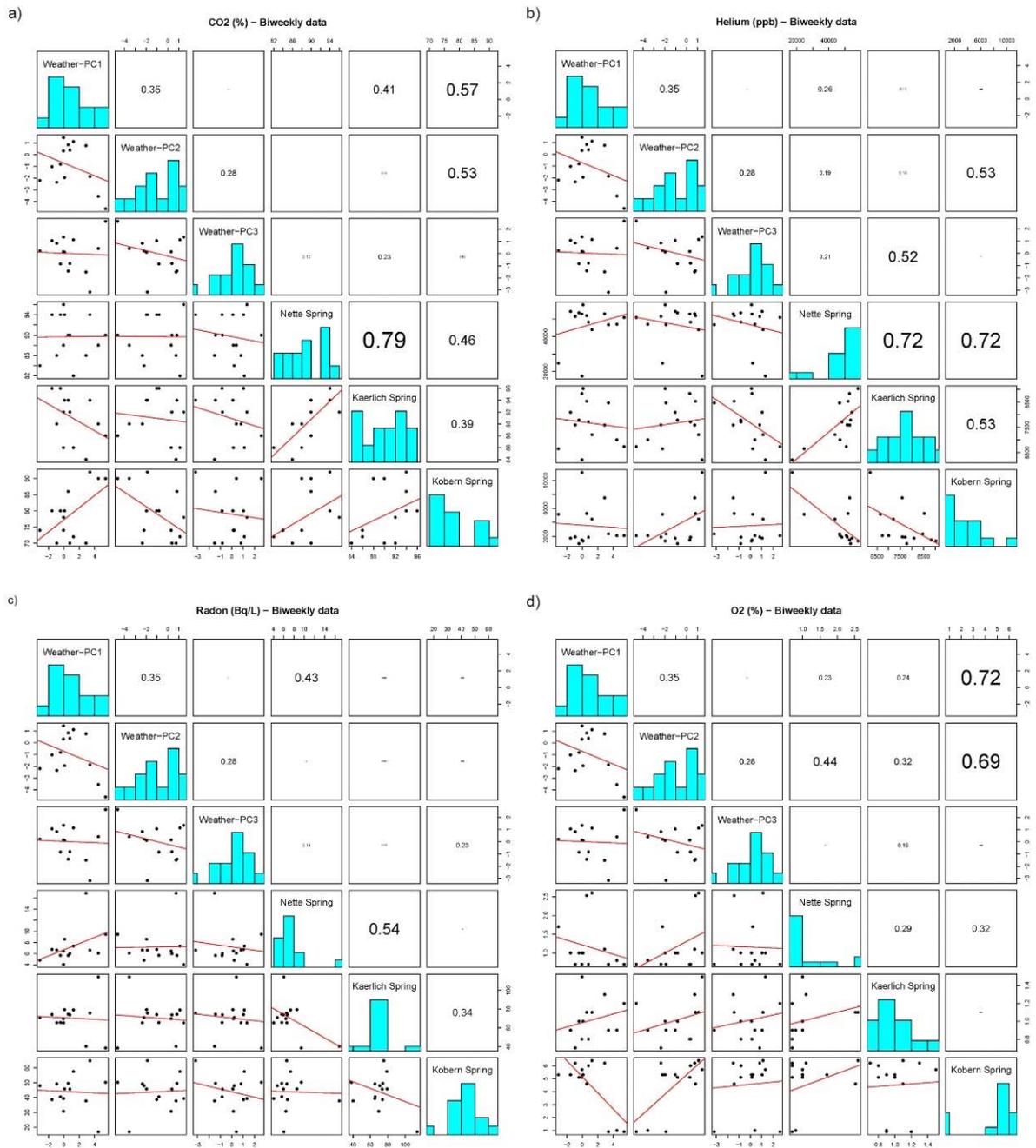

**Annex 2b    Correlation coefficients of the principal components analysis of weather variables with geogenic gas concentrations for (a) CO₂, (b) He, (c) Rn and (d) O₂ for the three mineral springs during the 4-W campaign**

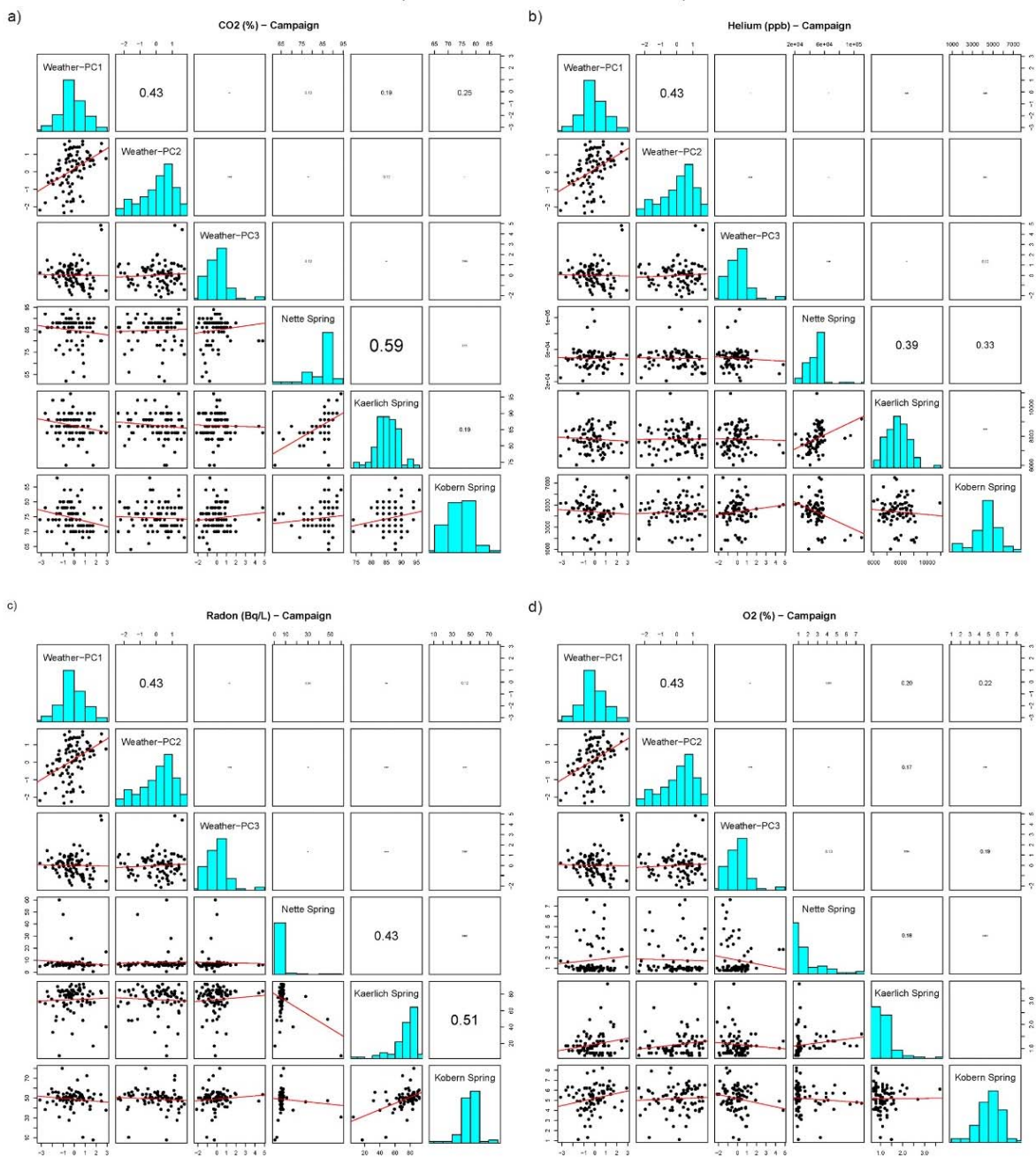

4-W (8-hours; 12.07. – 11.08.2016)

**Annex 3  Correlation coefficients between fluctuation patterns of geogas concentrations (centered and scaled data) during the bi-weekly (a) and 8-hours (b) survey in the three mineral springs and weather conditions**

| (a) 7 M | Nette | Kärlich | Kobern | T (°C) | AP (hPa) | H (%) | WSpd (km/h) |
|---|---|---|---|---|---|---|---|
| **CO$_2$ (Vol.%)** | | | | | | | |
| Nette | 1 | | | | | | |
| Kärlich | **0.92** | 1 | | | | | |
| Kobern | 0.35 | 0.21 | 1 | | | | |
| T (°C) | 0.25 | 0.38 | **-0.64** | 1 | | | |
| AP (hPa) | 0.35 | 0.43 | -0.06 | 0.13 | 1 | | |
| H (%) | -0.13 | -0.20 | -0.22 | 0.02 | -0.43 | 1 | |
| WSpd (km/h) | 0.18 | 0.08 | **0.86** | **-0.58** | -0.05 | -0.49 | 1 |
| **He (ppb)** | | | | | | | |
| Nette | 1 | | | | | | |
| Kärlich | **0.75** | 1 | | | | | |
| Kobern | **-0.59** | **-0.53** | 1 | | | | |
| T (°C) | 0.46 | 0.39 | -0.20 | 1 | | | |
| AP (hPa) | 0.24 | 0.42 | 0.05 | 0.13 | 1 | | |
| H (%) | -0.18 | 0.01 | -0.02 | 0.02 | -0.43 | 1 | |
| WSpd (km/h) | 0.21 | 0.08 | -0.18 | **-0.58** | -0.05 | -0.49 | 1 |
| **Rn (Bq/L)** | | | | | | | |
| Nette | 1 | | | | | | |
| Kärlich | 0.04 | 1 | | | | | |
| Kobern | 0.09 | 0.42 | 1 | | | | |
| T (°C) | 0.10 | -0.38 | **-0.71** | 1 | | | |
| AP (hPa) | 0.06 | **-0.66** | 0.01 | 0.13 | 1 | | |
| H (%) | -0.28 | 0.28 | -0.11 | 0.02 | -0.43 | 1 | |
| WSpd (km/h) | 0.13 | 0.31 | 0.45 | **-0.58** | -0.05 | -0.49 | 1 |
| **O2 (Vol.%)** | | | | | | | |
| Nette | 1 | | | | | | |
| Kärlich | 0.34 | 1 | | | | | |
| Kobern | 0.40 | -0.09 | 1 | | | | |
| T (°C) | 0.19 | -0.09 | **0.80** | 1 | | | |
| AP (hPa) | 0.06 | -0.47 | 0.18 | 0.13 | 1 | | |
| H (%) | -0.11 | 0.50 | 0.20 | 0.02 | -0.43 | 1 | |
| WSpd (km/h) | -0.26 | -0.04 | **-0.85** | **-0.58** | -0.05 | -0.49 | 1 |

| (b) 4-W | Nette | Kärlich | Kobern | T (°C) | AP (hPa) | H (%) | WSpd (km/h) |
|---|---|---|---|---|---|---|---|
| **CO$_2$ (Vol.%)** | | | | | | | |
| Nette | 1 | | | | | | |
| Kärlich | **0.65** | 1 | | | | | |
| Kobern | 0.12 | 0.35 | 1 | | | | |
| T (°C) | 0.25 | 0.24 | **0.63** | 1 | | | |
| AP (hPa) | -0.14 | 0.03 | -0.13 | -0.15 | 1 | | |
| H (%) | -0.05 | -0.28 | -0.36 | -0.43 | -0.17 | 1 | |
| WSpd (km/h) | 0.00 | -0.09 | **-0.53** | **-0.74** | 0.03 | **0.50** | 1 |
| **He (ppb)** | | | | | | | |
| Nette | 1 | | | | | | |
| Kärlich | **0.76** | 1 | | | | | |
| Kobern | 0.12 | 0.00 | 1 | | | | |
| T (°C) | 0.24 | 0.27 | **0.65** | 1 | | | |
| AP (hPa) | -0.16 | -0.32 | 0.05 | -0.15 | 1 | | |
| H (%) | -0.26 | -0.19 | -0.24 | -0.43 | -0.17 | 1 | |
| WSpd (km/h) | -0.02 | -0.13 | -0.34 | **-0.74** | 0.03 | **0.50** | 1 |
| **Rn (Bq/L)** | | | | | | | |
| Nette | 1 | | | | | | |
| Kärlich | 0.44 | 1 | | | | | |
| Kobern | 0.19 | **0.52** | 1 | | | | |
| T (°C) | 0.07 | -0.16 | 0.07 | 1 | | | |
| AP (hPa) | -0.31 | -0.24 | -0.19 | -0.15 | 1 | | |
| H (%) | -0.10 | -0.14 | -0.21 | -0.43 | -0.17 | 1 | |
| WSpd (km/h) | -0.12 | 0.04 | -0.17 | **-0.74** | 0.03 | **0.50** | 1 |
| **O$_2$ (Vol.%)** | | | | | | | |
| Nette | 1 | | | | | | |
| Kärlich | **0.69** | 1 | | | | | |
| Kobern | -0.13 | 0.00 | 1 | | | | |
| T (°C) | -0.19 | -0.07 | -0.48 | 1 | | | |
| AP (hPa) | 0.15 | 0.20 | 0.23 | -0.15 | 1 | | |
| H (%) | 0.08 | 0.30 | 0.13 | -0.43 | -0.17 | 1 | |
| WSpd (km/h) | -0.03 | 0.07 | 0.47 | **-0.74** | 0.03 | **0.50** | 1 |

T = Temperature; AP = Air Pressure; WSpd = Windspeed

**Annex 4      Correlations between fluctuation patterns of geogases (CO₂, He, Rn, O₂) and earth tides for the 4-W sampling**

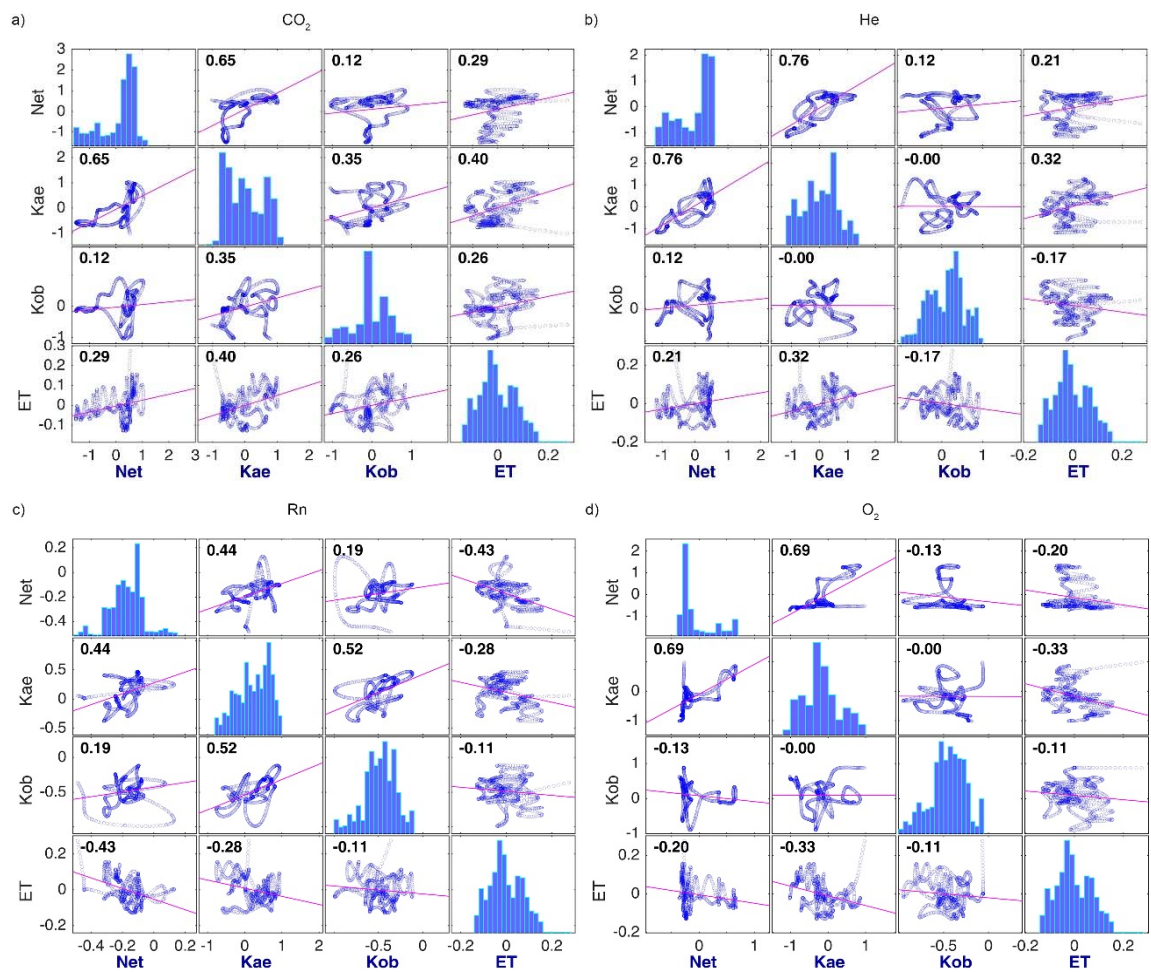